\documentclass[preprint]{aastex}

\newcommand \Msol {M_{\odot}}
\newcommand \Rsol {R_{\odot}}
\newcommand \eb {EROS BLG-2000-5}
\newcommand \VmI {($V$-$I$)$_0$}

\newcommand \Teff {T_{\rm eff}}
\newcommand \chimin {\chi^2_{\rm min}}
\newcommand \Kel {\,{\rm K}}
\newcommand \mn {\,\mu{\rm m}}
\newcommand \dchi {\Delta\chi^2}

\begin{document}
\title{
High-Precision Limb-Darkening Measurement of a K3 Giant Using Microlensing
}
\author{Dale L. Fields\altaffilmark{1}}
\author{and}
\author{
M. D. Albrow\altaffilmark{2,3},
J. An\altaffilmark{1,4},
J.-P. Beaulieu\altaffilmark{5},
J. A. R. Caldwell\altaffilmark{6},
D. L. DePoy\altaffilmark{1},\\
M. Dominik\altaffilmark{7},
B. S. Gaudi\altaffilmark{8,9},
A. Gould\altaffilmark{1},
J. Greenhill\altaffilmark{10},
K. Hill\altaffilmark{10},
U. G. J\o rgensen\altaffilmark{11},\\
S. Kane\altaffilmark{7},
R. Martin\altaffilmark{12},
J. Menzies\altaffilmark{6},
R. W. Pogge\altaffilmark{1},
K. R. Pollard\altaffilmark{2},
P. D. Sackett\altaffilmark{13},\\
K. C. Sahu\altaffilmark{3},
P. Vermaak\altaffilmark{6},
R. Watson\altaffilmark{10}, and
A. Williams\altaffilmark{12}
}\author{(The PLANET Collaboration)}
\author{and}
\author{
J.-F. Glicenstein\altaffilmark{14},
and
P. H. Hauschildt\altaffilmark{15}
}

\altaffiltext{1}
{Department of Astronomy, the Ohio State University,
140 West 18th Avenue, Columbus, OH 43210, USA}
\altaffiltext{2}
{Department of Physics \& Astronomy, University of Canterbury,
Private Bag 4800, Christchurch, New Zealand}
\altaffiltext{3}
{Space Telescope Science Institute,
3700 San Martin Drive, Baltimore, MD 21218, USA}
\altaffiltext{4}
{Institute of Astronomy, University of Cambridge,
Madingley Road, Cambridge CB3 OHA, United Kingdom}
\altaffiltext{5}
{Institut d'Astrophysique de Paris, INSU-CNRS,
98 bis Boulevard Arago, F 75014 Paris, France}
\altaffiltext{6}
{South African Astronomical Observatory,
P.O. Box 9, Observatory, 7935 South Africa}
\altaffiltext{7}
{School of Physics \& Astronomy, University of St.~Andrews,
North Haugh, St.~Andrews, Fife KY16 9SS, UK}
\altaffiltext{8}
{School of Natural Sciences, Institute for Advanced Study,
Einstein Drive, Princeton, NJ 08540, USA}
\altaffiltext{9}{Hubble Fellow}
\altaffiltext{10}
{Physics Department, University of Tasmania,
G.P.O. 252C, Hobart, Tasmania 7001, Australia}
\altaffiltext{11}
{Niels Bohr Institute, Astronomical Observatory,
Juliane Maries Vej 30, 2100 Copenhagen, Denmark}
\altaffiltext{12}
{Perth Observatory,
Walnut Road, Bickley, Western Australia 6076, Australia}
\altaffiltext{13}
{Research School of Astronomy \& Astrophysics, Australian National
University, Mt Stromlo, Weston ACT 2611, Australia.}
\altaffiltext{14}
{DAPNIA-SPP, CE-Saclay, F-91191 Gif-sur-Yvette, France}
\altaffiltext{15}
{Universit\"at Hamburg, Hamburger Sternwarte, Gojenbergsweg 112, D-21029
Hamburg, Germany}

\begin{abstract}

We obtain high-precision limb-darkening measurements in five bands ($V$,
$V_E$, $I_E$, $I$, and $H$) for the K3 III ($\Teff=4200\,{\rm K}$,
[Fe/H]$=+0.3$, $\log g=2.3$) source of the Galactic bulge microlensing
event \eb. These measurements are inconsistent with the predictions of
atmospheric models at $>10\,\sigma$.  While the disagreement is present
in all bands, it is most apparent in $I$, $I_E$ and $V_E$, in part
because the data are better and in part because the intrinsic
disagreement is stronger.  We find that when limb-darkening profiles are
normalized to have unit total flux, the $I$-band models for a broad
range of temperatures all cross each other at a common point.  The solar
profile also passes through this point.  However, the profile as
measured by microlensing does not.  We conjecture that the models have
incorporated some aspect of solar physics that is not shared by giant
atmospheres.

\end{abstract}

\section{Introduction} \label{sec:intro}

The brightness profiles (limb darkenings) of stars are a potentially
powerful probe of their atmospheres as a function of depth.  At each
point along the projected radius of a star, the observed flux originates
from a range of physical depths, the deepest of which (the surface of
last scattering) increases with physical radius as one's line of sight
progresses from the center of the star towards its limb.  Hence, since
stellar temperatures generally fall towards the surface, one expects
that the limb will appear cooler (and therefore redder and fainter) than
the center.  If model atmospheres accurately reflect the physical
conditions of the star as a function of depth, they should reproduce the
star's limb-darkening profile.

Because limb darkening is a photometric quantity, it can, in principle,
be measured to high precision.  The drawback is that one has to be able
to determine where on the star the light is coming from.  Historically,
there are two ways this is done.  The first is resolving the star.  The
most obvious example would be the Sun \citep{PW}.  Recent advances in
interferometry have allowed one to resolve the surfaces of the highest
angular-diameter stars \citep{Betel} and have in recent years provided
data good enough to begin challenging models.  The second method is by
occultation, either by an object in our solar system or one orbiting the
observed star.  The Moon is the only occulter used in our solar system.
However, while lunar occultations are sufficiently precise to
demonstrate that limb-darkened models are superior to uniform-brightness
models, they lack the precision to test limb-darkening models.  If the
occulting body is in the source's system, it can be a star or a planet.
If it is a star, the system is more properly referred to as an eclipsing
binary.  While such systems would seem to have great potential, it is
extremely difficult to disentangle the limb-darkened profile from other
parameters describing the fit to an eclipsing binary lightcurve
\citep{P84, P85}.  The first extra-solar transiting planet to be
discovered is HD 209458b \citep{char00}.  Because the planet is much
smaller and darker than the star, its transits can be used to trace the
stellar light profile in great detail \citep{Br01}.  Moreover, it is
generally expected that ongoing and future transit surveys will turn up
several more such systems.

There is one other method that can distinguish between light coming from
different parts of a star.  If a star passes through a microlensing
pattern, different parts of the star are magnified by different amounts.
In practice, differential magnification is significant only when the
star passes through a caustic, which is a region of formally infinite
magnification for a point source.  Up to now, the best measurement of
microlens limb darkening has been from MACHO 97-BLG-28 \citep{PL_SMC}.
This is because the event included a cusp crossing and thus the
magnification pattern was sharper, thereby giving better resolution
across the star.  In addition, the data for this event are very good.
\citet{PL_SMC} were able to measure two limb-darkening parameters each
in $V$ and $I$ for the K giant source.  They demonstrated that the
resulting surface profiles are in reasonable agreement with the
predictions of atmospheric models for stars of the same spectral type.
However, they were not able to challenge these models.

\citet{SMC} obtained a linear limb-darkening coefficient in each of four
bands for an A star in the Small Magellanic Cloud (MACHO 98-SMC-1).
They confirmed the expected trend of increased limb darkening toward the
blue, but the measurements are not precise enough to challenge models.
The primary difficulty is that the source star is extremely faint,
$I\sim22$, so that even when it is highly magnified, the signal-to-noise
ratio (S/N) is modest.

\citet{PL_MB9741} obtained linear limb-darkening coefficients for a red
clump giant (MACHO 97-BLG-41).  Even though the event itself was quite
favorable, with three caustic crossings and a cusp crossing, bad weather
and bad luck combined to limit the sensitivity of the data to limb
darkening.

The limb darkening analysis of OGLE-1999-BUL-23 by \citet{PL_OB9923} was
a major breakthrough in this subject.  They developed a method to
simultaneously compare limb-darkening measurements in two bands ($V$ and
$I$) to the predictions of a whole suite of atmospheric models.  The
analysis demonstrated a conflict only at the $2\,\sigma$ level, so no
significant conclusions could be drawn.  However, if the density of
measurements had been higher or the errors smaller, this technique would
have been able to give observational input into atmospheric modeling of
the limb darkening of a moderately evolved star for the first time.

The binary-lens microlensing event, \eb\ provides the best constraints
on limb darkening by any microlensing event.  This event has a caustic
crossing that is four days long.  This extraordinarily long timescale
and the generally excellent weather for all four days at all five
observatories combine to yield an extremely high density of coverage of
the source crossing in units of its own radius.  The first caustic
crossing is well measured (fortuitously since its onset cannot be
predicted beforehand), and the event contains a cusp approach in
addition to the two caustics.  Hence, this event is better constrained
than any other microlensing event \citep{A02}.  With the physical
parameters of the event well constrained, higher order terms in the
microlensing parameterization (such as limb darkening) can be precisely
determined.  Finally, we have spectroscopic data for this star (taken
when it was highly magnified) that give us independent information about
its temperature, metallicity and surface gravity.  This combination of
information allows us to to confront model atmospheres of what we
determine to be a K3 III star in all five bands.  We will test the
Kurucz ATLAS models \citep{Cl00} in the Johnson-Cousins $V$, $I$ and $H$
bands, and the Hauschildt Next$^2$Gen models in the same Johnson-Cousins
bands plus two EROS bands, $V_E$ and $I_E$.  The Next$^2$Gen models are
the current versions of the NextGen models described in
\citet{HDwarf,HGiant}.

\section{Data} \label{sec:data}

In our analysis of \eb\ we make use of 11 data sets in five filters.
The PLANET collaboration contributes nine data sets in three standard
filters, 3 sets in $V$, 4 in $I$ and 2 in $H$.  A description of the $I$
band data can be found in \citet{A02}. The $V$ band data are very
similar in quality to the $I$ band, the main difference being that they
contain about half the number of points.  The $H$ band data were taken
at SAAO and YALO by the instruments DANDICAM and ANDICAM, respectively.
The instruments and procedures are identical in these two cases, and
each contains a Tek $2048 \times 2048$ CCD and a Rockwell $1024 \times
1024$ HgCdTe IR array. The light path contains a dichroic that allows
optical and near-IR images to be obtained simultaneously at the same
position on the sky.  The $H$ band images are constructed by averaging
five contiguous dithered frames of 60s each. The dithered images are
flat-field corrected and then used to create a median sky image, which
is subtracted from the individual frames before they are shifted and
co-added.  The last two data sets are from the EROS collaboration.
These consist of $V$-like and $I$-like bands ($V_E$ and $I_E$) that are
described in \citet{EROS_SPEC}.  In the same manner as \citet{A02}, each
data set is cleaned of bad points and has its errors rescaled so that
the reduced $\chi^2$ is unity at the best-fit model.  Note that the
number of points in the PLANET $I$ band data sets may be individually
different from that given in \citet{A02} as these cuts were done
independently.  However, since the total difference is only 10 points
out of 1287, the impact on our conclusions is negligible.  The
attributes of the data sets are given in Table~\ref{tab:data}.

\section{Model} \label{sec:model}

We continue the model formalism of \citet{A02}, which contains 11
geometric parameters.  Seven of these are static binary lens parameters:
the lens separation in units of the Einstein radius $d_{t_c}$
(henceforth simply $d$), the binary lens mass ratio $q$, the angle
between the direction of motion of the source and the binary lens axis
$\alpha^{\prime}$, the distance between the cusp and the source at
closest approach $u_c$, the time taken to travel an Einstein radius
$t^{\prime}_E$, the time of closest approach to the cusp $t_c$, the
ratio of source radius to Einstein radius $\rho_*$.  Two are rotational
parameters: $\dot{d}$ and $\omega$, and two are vector components of
microlens parallax: $\pi_{E,\parallel}$ and $\pi_{E,\perp}$.  The
derivation of this parameterization and its relation to the standard
formalism is given in \citet{A02}.  Our geometrical solution and that of
\citet{A02} are given in Table~\ref{tab:soln}.

In addition, each observatory and band has its own five photometric
parameters: the unmagnified source flux $f_s$, the blend flux $f_b$, a
linear seeing correction $\eta_s$, a linear limb-darkening parameter
$\Gamma$, and a square-root limb-darkening parameter $\Lambda$.  These
photometric parameters are returned from a linear fit to the
magnification curve determined by the 11 geometrical parameters.  For
each band ($V$, $I$, and $H$) that is observed from several
observatories, all observatories are constrained to give the same values
of $\Gamma$ and $\Lambda$.

The form of the limb-darkening law we use is,
\begin{equation}
\label{eqn:limblaw}
S_\lambda (\vartheta)
=
\bar{S}_\lambda
\left[\left(1-\Gamma_\lambda-\Lambda_\lambda\right)
+\frac{3\Gamma_\lambda}{2}\cos\vartheta+
\frac{5\Lambda_\lambda}{4}\sqrt{\cos\vartheta}\right]
\,,
\end{equation}
which conserves flux independent of $\Gamma$ and $\Lambda$, with
$\bar{S}_\lambda$ being the mean surface brightness of the source and
$\vartheta$ the angle between the normal to the stellar surface and the
line of sight.  This law is a different form of the more widely-used,
\begin{equation}
\label{eqn:slimb}
S_\lambda (\vartheta)
=
S_\lambda (0)
\left[1-c_\lambda(1-\cos\vartheta)
-d_\lambda(1-\sqrt{\cos\vartheta})\right]
\ .
\end{equation}
It should be noted, however, that equation~(\ref{eqn:slimb}) is normalized
to the flux at the center, and thus the total flux is a function of
$S_\lambda (0)$, $c_\lambda$ and $d_\lambda$.

The transformation of the coefficients in equation~(\ref{eqn:limblaw})
to the usual coefficients used in equation~(\ref{eqn:slimb}) is given by,
\begin{equation}
\label{eqn:cd}
c_\lambda=\frac{6\Gamma_\lambda}{4+2\Gamma_\lambda+\Lambda_\lambda}
\,,\ \ \
d_\lambda=\frac{5\Lambda_\lambda}{4+2\Gamma_\lambda+\Lambda_\lambda}
\,\end{equation}
while the inverse transformation is
$\Gamma_{\lambda}=10c_{\lambda}/(15-5c_{\lambda}-3d_{\lambda})$ and
$\Lambda_{\lambda}=12d_{\lambda}/(15-5c_{\lambda}-3d_{\lambda})$.  The
limb-darkening parameters are primarily determined by the behavior of
the light curve between the time the source edge enters the caustic and
the time the source center enters the caustic (and the inverse of this
process as the source leaves the caustic).

We then take this model and expand it to include all five bands,
constraining all observations in the same band to give the same
limb-darkening parameters.  However, since there are no seeing data for
EROS, these two bands do not have seeing corrections.  This gives our
model a total of 11 (geometric) + [5 (photometric) $\times$\ 11 (data
sets) $-2$ (EROS seeing coefficients)] = 64 fit parameters, which are
then subject to [(2 $V$ + 3 $I$ + 1 $H$ )$\times2$(limb-darkening
parameters)]$=12$ constraints.

\section{Analysis} \label{sec:analysis}

\subsection{$\chi^2$ Minimization} \label{sec:grid}

We then take this model and search for minima in $\chi^2$ in the
11-dimensional space of the geometrical parameters.  We employ a simple
grid search algorithm and not a more efficient technique such as simplex
because the (apparent) $\chi^2$ surface is rough with many false minima.
This problem forces us to restrict our automated grid search to nine of
the geometrical parameters and to step through the remaining two ($d$,$q$)
``by hand.''

Even so, we find that we can locate the true minimum only to within
$\sim0.02$ in $d$ and $q$, despite the fact that the true errors in
these parameters (as determined from the curvature of the $\chi^2$
surface measured over larger scales) are $<0.01$.  That is, the apparent
$\chi^2$ varies by $\sim10$ for the same ($d$,$q$) when we initialize
our search using different values of the other nine parameters.  We
believe that this roughness is most likely due to numerical noise,
rather than roughness in the ``true'' $\chi^2$ surface.  However,
regardless of the exact cause of the roughness of the surface, its
impact via increased uncertainty in ($d$,$q$) on the errors in the
limb-darkening parameters must be assessed.  This will be done in
\S\ref{sec:errors}.

We start the grid search near the minimum found by \citet{A02}.  To save
computation cycles, we first do a grid search using only $I$, $I_E$, and
$H$ band data.  After the new minimum has been approximately located,
the $V$ and $V_E$ band data are finally used.

\begin{figure}
\plotone{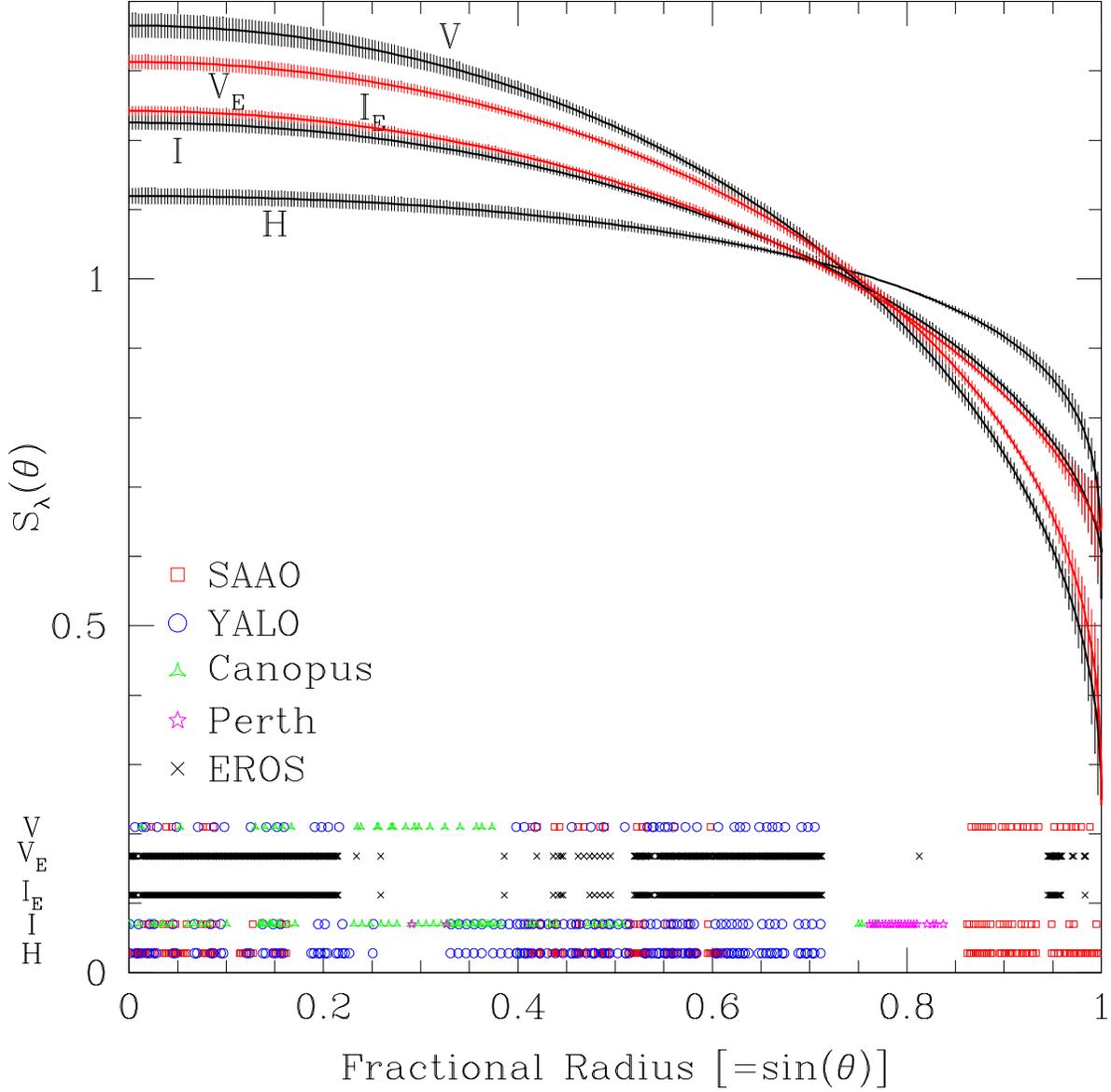}
\caption{\label{fig:fivebands}
The derived brightness profiles for the source star of \eb\ in all five
bands, $V$, $V_E$, $I_E$, $I$ and $H$.  The shaded regions around the
curves are the $3\,\sigma$ error envelopes.  Also shown are points
indicating when data were taken at each observatory, expressed in
distance from the caustic to the center of the source in units of the
source radius.  At this point, the magnification pattern over the source
profile is discontinuous, which gives us precise information about the
profile at that point.  For example, since the $I$ band has almost
continuous coverage across the entire source star, we can be confident
that the brightness profile well represents reality.  The $V$ and $H$
bands have a gap in coverage around $\sin \theta=0.8$ and thus we can be
less confident of the profile there.  The EROS data, produced by a
single observatory, has large gaps, but are still able to make the two
parameter fit on account of their very dense coverage when the
observatory was active.  }\end{figure}

As stated above, the fitting routine finds both linear $\Gamma$ and
square-root $\Lambda$ limb-darkening parameters.  However, in the case
of the $H$ band, the degeneracy between $\Gamma_H$ and $\Lambda_H$ is
too severe for this 2-parameter fit to provide useful information.
Because $\Gamma_H$ is consistent with zero, we constrain it to be
exactly zero and report only $\Lambda_H$.  The analysis is then redone
giving us 64 parameters subject to 13 constraints, which, since two
parameters are constrained to be zero, is equivalent to 62 parameters
with 11 constraints.  The limb darkening of the best-fit microlens model
is shown in Figure~\ref{fig:fivebands}.

\subsection{Errors} \label{sec:errors}

\begin{figure}
\plotone{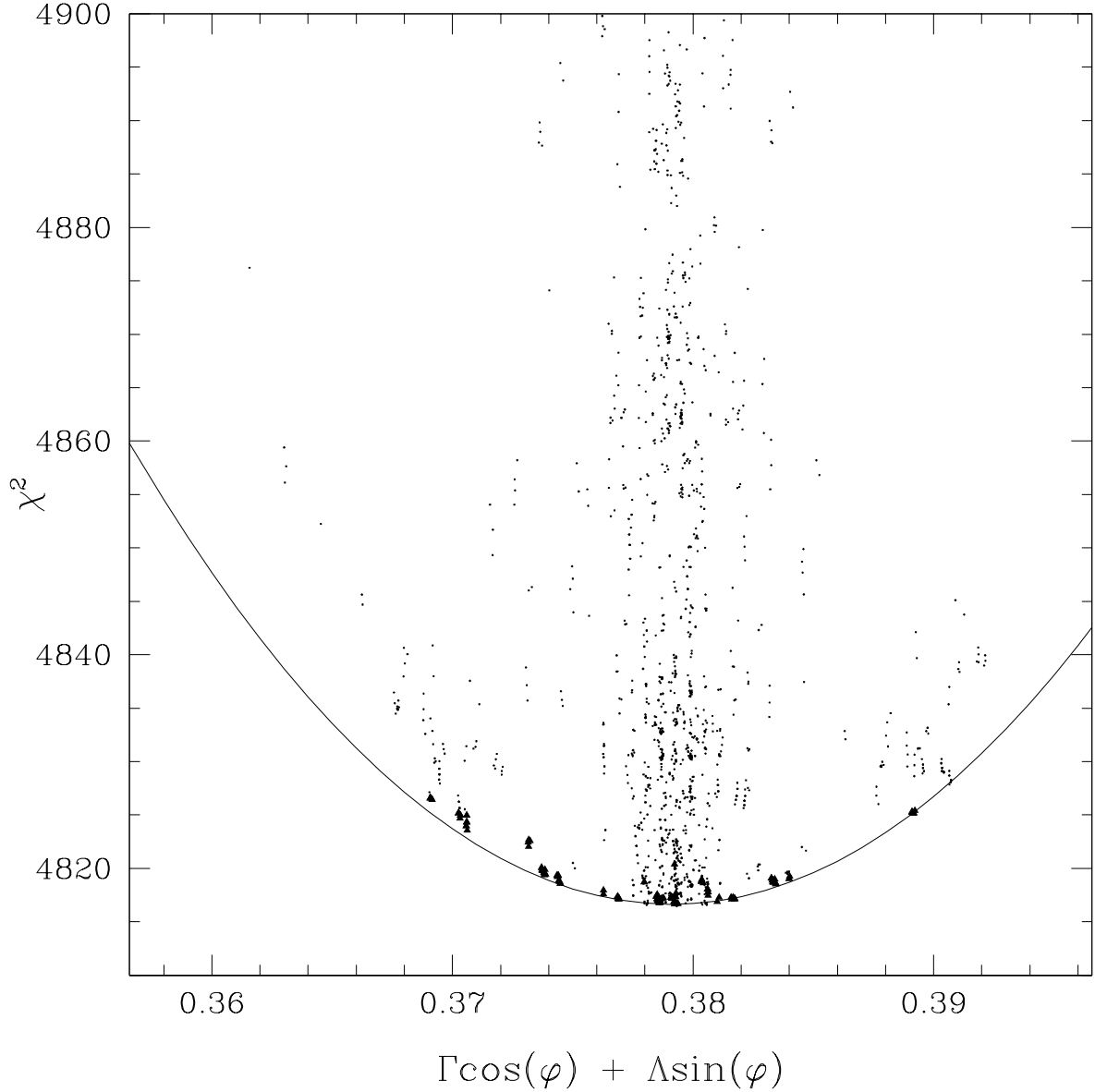}
\caption{\label{fig:parabola}
The $I$ band $\chi^2$ surface plotted as ($\Gamma$, $\Lambda$, $\chi^2$)
and viewed along the minor axis of the error ellipse.  Each point
represents a slightly varied geometry and its associated $\chi^2$ and
limb-darkening parameters.  The filled triangles are those points we use
to create the covariance matrix.  The parabola is the slice through the
geometric-error paraboloid for this particular viewing angle.  Numerical
noise will lift points in $\chi^2$ at a particular ($\Gamma$,$\Lambda$),
and is responsible for the larger $\chi^2$ among those points that we do
not use.  }\end{figure}

Contributions to the error in the limb-darkening parameters can be
broken down into three sources: the photometric (which include
limb-darkening) parameters, the geometric parameters we minimize over,
and $d$ and $q$.  The covariance matrix from the photometric parameters
is easily obtained as it is a byproduct of the linear fit that solves
for the photometric parameters.  This is done holding all the geometric
parameters fixed.  We then apply the hybrid statistical error analysis
as given in Appendix D of \citet{A02} to determine the combined
covariance matrix for the nine geometrical + 53 photometric parameters.
While this approach does yield an invertible covariance matrix, there
are reasons not to trust this method in this particular instance.  We
have stated above in \S\ref{sec:grid} that the apparent $\chi^2$ surface
is rough, with numerical noise a possible culprit.  An effect of
numerical noise upon the ideal $\chi^2$ parabola is to artificially
raise the $\chi^2$ at any particular point in parameter space.  This is
illustrated in Figure~\ref{fig:parabola}.  A good approximation to the
true (numerical-noise free) parabola can be found by fitting to the
envelope as noise will not decrease $\chi^2$.  We fit for the error
induced by the nine geometric parameters that we minimize over in this
fashion.  We fit for the error in each band separately.  In all cases,
we find that the $\Gamma$-$\Lambda$ error ellipse induced by the nine
geometrical parameters is small compared to that induced by the
photometric parameters.  The two sets of error ellipses are
well-aligned, differing in direction by only a few degrees.  We add the
resulting covariance matrices to obtain the geometric $+$ photometric
errors.

Up to this point, we have not yet taken into account the error induced
by $d$ and $q$.  The roughness of the $\chi^2$ surface is also a factor
in this analysis.  Unfortunately, while it is theoretically possible to
use the same method that we use on the other geometrical parameters, it
would take several orders of magnitude more computational resources than
we currently have to properly populate the geometrical figure analogous
to Figure~\ref{fig:parabola}.  We instead investigate whether and to
what extent the additional error in our limb-darkening parameters
induced by $d$ and $q$ will affect our conclusions.

To this end, we first define
\begin{equation}
\label{eqn:chi2}
(\dchi)_{LD}=\sum_{i,j}{\delta}{a_i}{b_{ij}}{\delta}{a_j},
\end{equation}
where ${\delta}a_i$ is a vector whose components are the differences
between the limb-darkening parameters of our best-fit model and a
comparison model.  Here, $b_{ij}\equiv\,(c_{ij})^{-1}$ and $c_{ij}$ is
the covariance matrix of the limb-darkening parameters evaluated at the
best-fit model.  We define equation~(\ref{eqn:chi2}) as $\dchi$ because
it is the distance, expressed in $\chi^2$ and normalized by the error
ellipsoid, between models that both attempt to describe the data.  We
then compare our best-fit microlens model with other microlens models
over the roughly ($0.02\times0.02$) region of the ($d$,$q$) space over
which $\chi^2$ cannot be properly minimized on account of the roughness
in the $\chi^2$ surface described in \S\ref{sec:grid}.  We find that the
limb-darkening parameters vary by less than $1\,\sigma$.  By comparison,
as we will show in \S\ref{sec:atlas} and \S\ref{sec:ng}, the
limb-darkening parameters of our best-fit model differ from those
predicted by the stellar models by $>10\,\sigma$.  We conclude that the
additional errors resulting from both the uncertainty in $d$ and $q$ and
the underlying numerical noise do not affect our overall conclusions.
We recognize this additional error exists, but given that we have no way
to quantify it properly, we simply report the error induced by the
photometric and other nine geometric parameters in Table~\ref{tab:corr}.

\subsection{Independent Analysis of Source Star} \label{sec:indep}

When we compare our limb-darkening results with the atmospheric models
in \S\ref{sec:atlas} and \S\ref{sec:ng}, we wish to restrict
attention to models that are relevant to the source star.  We therefore
begin by summarizing the results of an analysis of the source's physical
properties as given by \citet{A02}.  Assuming no differential reddening
across the field, the dereddened color and magnitude of the source can
be found by measuring the source offset from the red clump identified in
a color-magnitude diagram (CMD) of the field: \VmI$=1.390\pm0.010$ and
$I_0=14.70\pm0.03$.  From its color and the fact that it is a giant (see
below), the source is a K3 III star with a corresponding
$\Teff=4200\Kel$.  The color and magnitude imply a source angular radius
$\theta_*=6.62\pm0.58\,\mu$as.  The source-lens relative proper motion
and source radial velocity imply that the source most probably lies in
the bulge, i.e., at $d_s\sim8$\,kpc, which implies a physical radius
$r_*=d_s\theta_*=11.4\pm1.0\,\Rsol$.  From kinematic information,
\citet{A02} find that while it is possible that the source lies in the
far disk (and so is physically bigger), it essentially cannot lie in the
near disk.  Since the source is either in the bulge, or in the far disk
$\sim500\,$pc above the plane, it must be a fairly old giant and
therefore have a mass $M\sim1\,\Msol$.  The mass and radius combine to
give $\log g\sim2.3$.  If it lies in the far disk at 12\,kpc then $\log
g$ can be as much as 0.3 dex smaller.

\citet{R03} have analyzed HIRES Keck spectra taken on the last two
nights of the caustic crossing, HJD 2451731.953 and HJD 2451732.950 when
the source center was approximately 0.25 and 0.75 source radii outside
the caustic.  By comparing these spectra with integrated spectra of
model sources, they estimate that $\Teff=4250\Kel$, $\log g=1.75$, and
[Fe/H]$=+0.29\pm0.04$ on the first night and that $\Teff=4450\Kel$,
$\log g=2.25$, and [Fe/H]$=+0.22\pm0.07$ on the second.  The gravity is
too loosely constrained in this analysis to be of any use, but
fortunately we have the photometric method applied to the \citet{A02}
results given above, which is both simpler and more robust.  Because the
source is differentially magnified while the models are not, this
spectroscopic approach is not fully self-consistent.  Nevertheless, we
expect the error induced to be modest, particularly on the first night
when the limb of the star is not particularly emphasized in the
integrated source light.  Since, in addition, both the observing
conditions and S/N were substantially better the first night, we adopt
$\Teff=4250\Kel$ and [Fe/H]$=+0.3$ as the spectroscopic determinations.
The former is in excellent agreement with the photometric determinations
described above.  We designate ($\Teff$,[Fe/H],$\log
g$)=(4200\,K,+0.3,2.3) as the most physical model (MPM).

If we are to use these estimates to define a viable region of model
atmosphere parameter space, we need error estimates as well.
Spectroscopic temperature estimates are routinely good to a hundred
Kelvins.  Similarly, our photometric temperature estimate can be off by
a hundred Kelvins depending on differential reddening.  As summarized
above, the source gravity is strongly constrained by the angular-size
measurement and distance estimates.  One could possibly push the source
into the near part of the bulge, or into the far disk, but that is all.
There is also some error associated with the mass estimation, but due to
age constraints the mass cannot be too far from $\sim1\,M_{\odot}$.  To
be safe, we budget a 50\% mass error.  Such errors give us $\pm0.38$ in
$\log g$.  The metallicity is the least well constrained.  This is only
a minor problem, for as we shall show in \S\ref{sec:atlas} and
\S\ref{sec:ng}, metallicity has only a minor effect upon the
model-atmosphere limb-darkening curves.  We set our lower metallicity
limit at solar.  Physically reasonable models (PRMs) would then be
$\Teff=4100$-$4300\Kel$, $\log g=1.9$-2.7 and [Fe/H]=0.0 - +0.3.

\subsection{Comparison with ATLAS Models} \label{sec:atlas}

\citet{Cl00} fits five different limb-darkening laws to a suite of ATLAS
model atmospheres supplied to him by R. Kurucz in 2000.  \citet{Cl00}
then reports the parameters for each of these limb-darkening laws.  We
compare our results to the linear + square root law, rather than the
favored four-parameter fit since we also use a linear + square root law
in our fitting and thus the coefficients are comparable.  Little is lost
by this substitution as the four-parameter and two-parameter fits differ
by much less than the difference between the microlensing-based and
atmospheric-model profiles.  We use equation~(\ref{eqn:chi2}) to create
our measure of goodness of fit, but because we are now comparing
theoretical atmospheric models to what we consider a parameterization of
reality, we term the result $\chi^2$.  Eq.~\ref{eqn:chi2} implicitly
assumes that the $\chi^2$ surface is parabolic which, since the
microlensing fit is non-linear, is not strictly the case.  However, as
shown in \S\ref{sec:errors}, the covariance matrix is dominated by the
linear part of the fit.  Hence the $\chi^2$ surface is nearly parabolic.
As before ${\delta}a_{i}$ is a difference in the limb-darkening
parameters, this time between an atmospheric model and our microlensing
result, and $b_{ij}$ is the inverse of $c_{ij}$, the covariance matrix
for the microlensing limb-darkening parameters.  We convert the
\citet{Cl00} $c_{\lambda}$ and $d_{\lambda}$ to $\Gamma_{\lambda}$ and
$\Lambda_{\lambda}$ using the inverse of equation~(\ref{eqn:cd}) to make
the comparison.  Since we have enforced $\Gamma_H\equiv0$ on the
microlensing models, we repeat this constraint on the atmosphere models.
To do this, we determine the ($\Gamma_H$,$\Lambda_H$) covariance matrix
by removing the constraint $\Gamma_H\equiv0$ in the microlensing fit.
We then use the method given in Appendix A of \citet{GA02} to enforce
the constraint on the
\citet{Cl00} data.  We restrict the comparison to the standard
(Johnson-Cousins) bands, $V$, $I$, and $H$, because \citet{Cl00} reports
limb-darkening parameters only for these.

Following the procedure pioneered by \citet{PL_OB9923}, we begin by
simultaneously comparing the microlensing limb-darkening parameters from
all three filters with the atmospheric-model parameters, taking full
account of the covariances among these five parameters.  We restrict our
investigation to those models with turbulent velocity $v_T=2$ only.
Models given by \citet{Cl00} with other $v_T$ do not span the full
parameter space required by our investigation, nor do we have
independent information that would distinguish among different $v_T$ as
we do for $\Teff$, [Fe/H] and $\log g$.  Special attention is paid to
two regions of the $\chi^2$ surface: the neighborhood of the MPM to
check for consistency between the atmospheric models and the microlens
data, and features around $\chimin$ (which may not be near the MPM) to
try to guide modelers in understanding the results of their simulations.
The ATLAS parameter grid does not contain the MPM, but the closest is
$\Teff=4250\Kel$, $\log g=2.5$ and [M/H]$=+0.3$.  We shall refer to this
as the MPM while within \S\ref{sec:atlas}.  Additionally, the ATLAS
models given in \citet{Cl00} have a larger grid spacing than the region
covered by the PRMs, so we investigate $\Teff=4000$-$4500\Kel$, $\log
g=2.0$ and $2.5$, and [M/H] from $0.0$ to $+0.3$.  We define consistency
as having a $\chi^2\lesssim4$.  We find that $\chi^2$ at the MPM is 164.
In fact, all of the PRMs are high, with the $\chi^2$ minimum at 45.
This in itself is a major concern.  That no model atmosphere agrees with
our data, regardless of its parameters is evident by the fact that the
$\chimin=37$.  This occurs at $\Teff=4500\Kel$, $\log g=3.0$ and $3.5$,
and [M/H]$=-0.3$, which is incompatible with the other evidence we have
about this star.  We find that, in all the ATLAS models in the vicinity
of the MPM, the differences in $\chi^2$ between $\log g=2.0$ and $2.5$
are small compared to those induced by changes in the other two
parameters.  This is true for the $I$ and $H$ bands as well, but $\log
g$ has more of an effect in the $V$ band.  Due to this lack of
distinguishability, we shall focus our investigation on $\log g=2.5$.
Any model whose gravity is not listed should be assumed to have $\log
g=2.5$.  To determine whether the large mismatch in limb-darkening
parameters comes primarily from one specific band, we investigate the
goodness of fit for each band separately.

\subsubsection{$V$ band} \label{sec:av}

\begin{figure}
\plotone{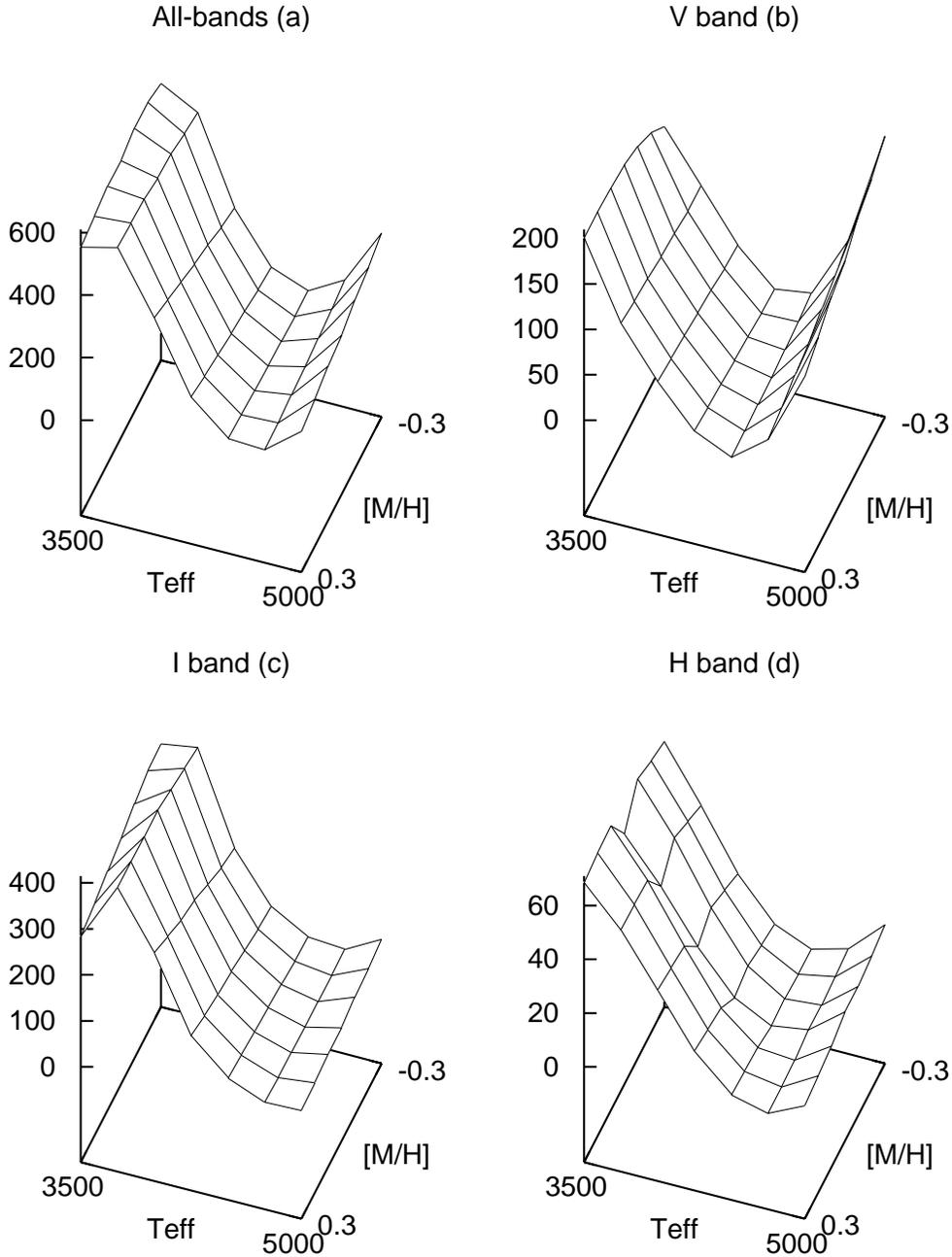}
\caption{\label{fig:chi2at}
Shown are the $\chi^2$ surfaces of the microlens-ATLAS comparison at
$\log g=2.5$ for (a) the $V$ + $I$ +$H$ bands combined, (b) the $V$
band, (c) the $I$ band and (d) the $H$ band.  The value of $\chi^2$ is
most dependent on $\Teff$, and less dependent on [M/H].  The dependence
on $\log g$ (not shown) is even weaker still.  The surface of every band
has a similar shape, a ``valley'' that runs through metallicity with
almost constant $\Teff$.  These structures mostly overlap, though the
$I$ band's is shifted to slightly higher $\Teff$.  }\end{figure}

We repeat our $\chi^2$ minimization over the space of ATLAS models
considering only the $V$ band parameters $\Gamma_V$ and $\Lambda_V$.
The MPM has a $\chi^2$ of 19.  We then look for PRMs that might be
consistent.  All points with $\Teff=4500\Kel$ have $\chi^2<4$, and are
thus consistent.  Within that consistency, higher [M/H] and $\log g$ are
favored.  Remember, however, that the grid spacing of the ATLAS models
is larger than the true permitted temperature range.  If we take this
into account, and note that at $\Teff=4250\Kel$, $\chi^2>11$, we must
downgrade $V$ band to marginal inconsistency.  The shape of the $\chi^2$
surface near $\chimin$ returns a ``valley'' (part of which is shown in
Fig.~\ref{fig:chi2at}) running along $\Teff=4500\Kel$.  Surface gravity
is more important in the $V$-band $\chi^2$ than in the sum over all the
bands.  Regardless, $\chi^2$ in the $V$ band alone still depends more
upon [M/H] than $\log g$.


\subsubsection{$H$ band} \label{sec:ah}

We perform an analysis for $H$ band in the same manner as in the
previous section.  The MPM has $\chi^2=16$, similar to that of the $V$
band.  The $H$ band also has consistency at the same level and manner as
$V$ band, at $\Teff=4500\Kel$ in the PRMs and the same caveat discussed
in \S\ref{sec:av} applies for the most part.  The $H$ band also has a
valley structure (Fig.~\ref{fig:chi2at}) in its $\chi^2$ surface
analogous to that in the $V$ band, though there is some dependence on
$\Teff$ in the range that we investigate.  This $\Teff$ dependence is
slight; the location of the``valley floor'' only shifts from
$\Teff=4500\Kel$ at [M/H]$=-0.3$ to $\Teff=4750\Kel$ at [M/H]$=+0.3$.
As in the all-bands, $\log g$ is essentially unimportant in the $H$
band.




\subsubsection{$I$ band} \label{sec:ai}

The $I$ band MPM has $\chi^2=128$.  None of its PRMs are consistent with
the microlens limb darkening; the lowest $\chi^2$ among them is 41.  The
$\chimin$ over the entire space of $I$ band models is still moderately
high at 14.  In general, it is the $I$ band that is causing most of the
discrepancy between models and data.  The atmospheric models that have
the lowest $\chi^2$ form a track in parameter space that varies smoothly
from solar metallicity dwarfs at $\Teff=4750\Kel$ to super metal-poor
supergiants at $\Teff=3500\Kel$.

At this point, we can ask whether the bands are consistent with each
other.  In a relative sense, they are, as the PRMs with the lowest
$\chi^2$ are always those with $\Teff=4500\Kel$, no matter the band.  We
defer discussion as to the possible causes of the disagreement between
the ATLAS atmospheric models and the microlens data until
\S\ref{sec:sources}, after we have investigated the Next$^2$Gen models.

\subsection{Comparison with Next$^2$Gen Models} \label{sec:ng}

We analyze the limb darkening of Next$^2$Gen models between
$\Teff=4000\Kel$ and $4600\Kel$ in $100\Kel$ increments, $\log g=0.0$
and $3.5$ in increments of $0.5$ and [Fe/H] of $-0.25$, $0.0$ and
$+0.3$.  The original format of these files is a spectrum between
3500\AA\ and $3\mn$ with a resolution of 0.5\AA\ at each of 99 points in
$\cos \theta$.
Having the full spectra enables us to create limb-darkening profiles in
non-standard bands.  We convolve the spectra with filter functions for
all five filters in the microlens data ($V$, $V_E$, $I_E$, $I$ and $H$).
We then use a simple linear fit to solve for the
($\Gamma_{\lambda}$,$\Lambda_{\lambda}$) for each filter except $H$ band
for which we fit for $\Lambda_{\lambda}$ alone.

\begin{figure}
\plotone{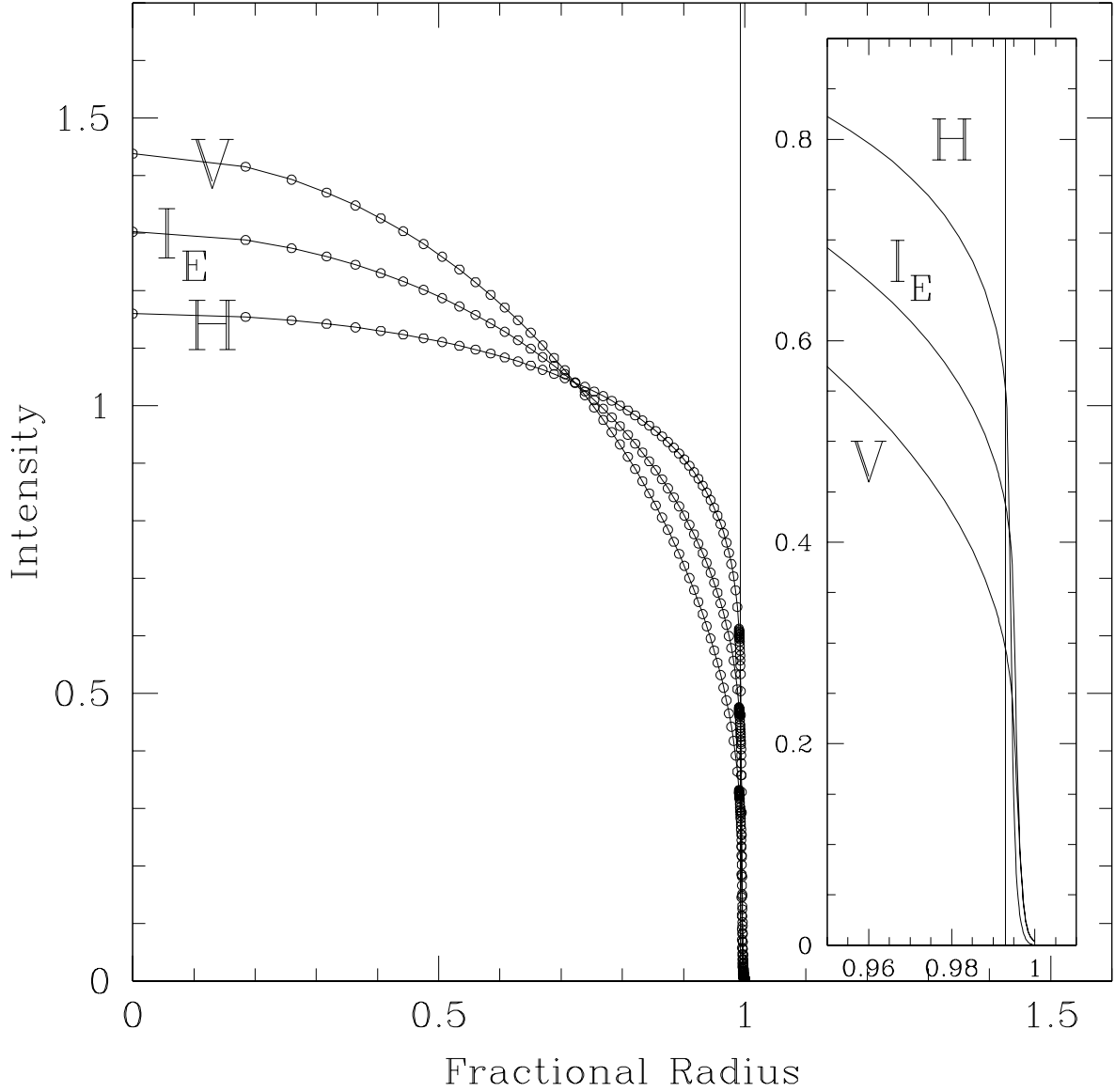}
\caption{\label{fig:nextgen}
Sample limb-darkening profiles from the Next$^2$Gen model corresponding to
the MPM ($\Teff=4200\Kel$, $\log g=2.5$, [Fe/H]$=+0.3$).  Shown are the
$V$, $I_E$ and $H$ bands.  Each circle is a radial point in the Next$^2$Gen
output.  Also shown is the cut in radius that we impose at $r=0.993$.
}\end{figure}

The primary difficulty in this procedure is the definition of the edge
of a star.  The Next$^2$Gen atmospheres have a steep drop off in
intensity whose location in radius varies with surface gravity.  Sample
profiles are shown in Figure~\ref{fig:nextgen}.  This feature cannot be
modeled by a linear + square root limb-darkening law, and because it
contributes almost nothing to the total flux, we decide to remove it.
This is further warranted because even if we had the formalism in our
microlensing code to fit this feature, we would not receive any useful
information since our sampling is not dense enough at the specific part
of the caustic exit during which the feature would be visible.  We
therefore excise this feature by removing all points outside of some
chosen radius.  We then rescale the value of the radius at each
remaining point by the factor necessary to set the outermost point's
radius equal to unity.  The radius is chosen by finding the point at which
the $H$ band (which should suffer the least amount of limb darkening),
drops steeply off.  This is something of a judgment call as individuals
will pick slightly different truncation radii.  This does not pose a
problem, however, as tests indicate that the Next$^2$Gen profile is
equally well fit by a two-parameter limb-darkening law out to radii
outside this steep drop off in flux (but not into the feature we are
removing).  In the profiles shown in Figure~\ref{fig:nextgen}, this
occurs around $r=0.995$, indicating that our cut of $r=0.993$ is
acceptable.  This procedure breaks down at low surface gravity.
Supergiants have such a small density gradient that the surface of last
scattering at different wavelengths varies greatly with radius.  This
would be a major concern for us if we did not have additional
information telling us that this star was a luminosity class III giant.
We use a separate truncation radius at each $\log g$, but not for each
wavelength.  Our adopted truncation radius varies between $r=0.88$ at
$\log g=0.0$ to $r=0.998$ at $\log g=3.5$.  Having a single truncation
radius for all bands would induce problems at the low $\log g$ end, but
such low surface gravities are already highly disfavored as discussed in
\S\ref{sec:indep}.

In performing the fit, we sample the profile at the radii corresponding
to the observations (see Fig.~\ref{fig:fivebands}), giving equal weight
to each point.  We evaluate the profile at these radii by interpolating
among the points given by the Next$^2$Gen model.  This produces a model
profile that is most weighted in the regions that are most densely
observed in the real data.

We then perform the same analysis as in \S\ref{sec:atlas}, creating a
$\chi^2$ surface between the microlens limb-darkening parameters and
those of the Next$^2$Gen models.  The only difference is that we now have
nine limb-darkening parameters instead of five because we are also
matching $V_E$ and $I_E$ in addition to $V$, $I$ and $H$.  As in the
previous section, no grid points in parameter space for the Next$^2$Gen
models coincide perfectly with the MPM, and the grid spacing does not
perfectly coincide with the range of the parameters covered by the PRMs.
In this section we will consider $\Teff=4200\Kel$, $\log g=2.5$ and
[Fe/H]$=+0.3$ as the MPM, and we will consider the range $\Teff=4100$ to
$4300\Kel$, $\log g=2.0$ and $2.5$, and [Fe/H]$=0.0$ to $+0.3$ as that
similar to our PRMs.

\begin{figure}
\plotone{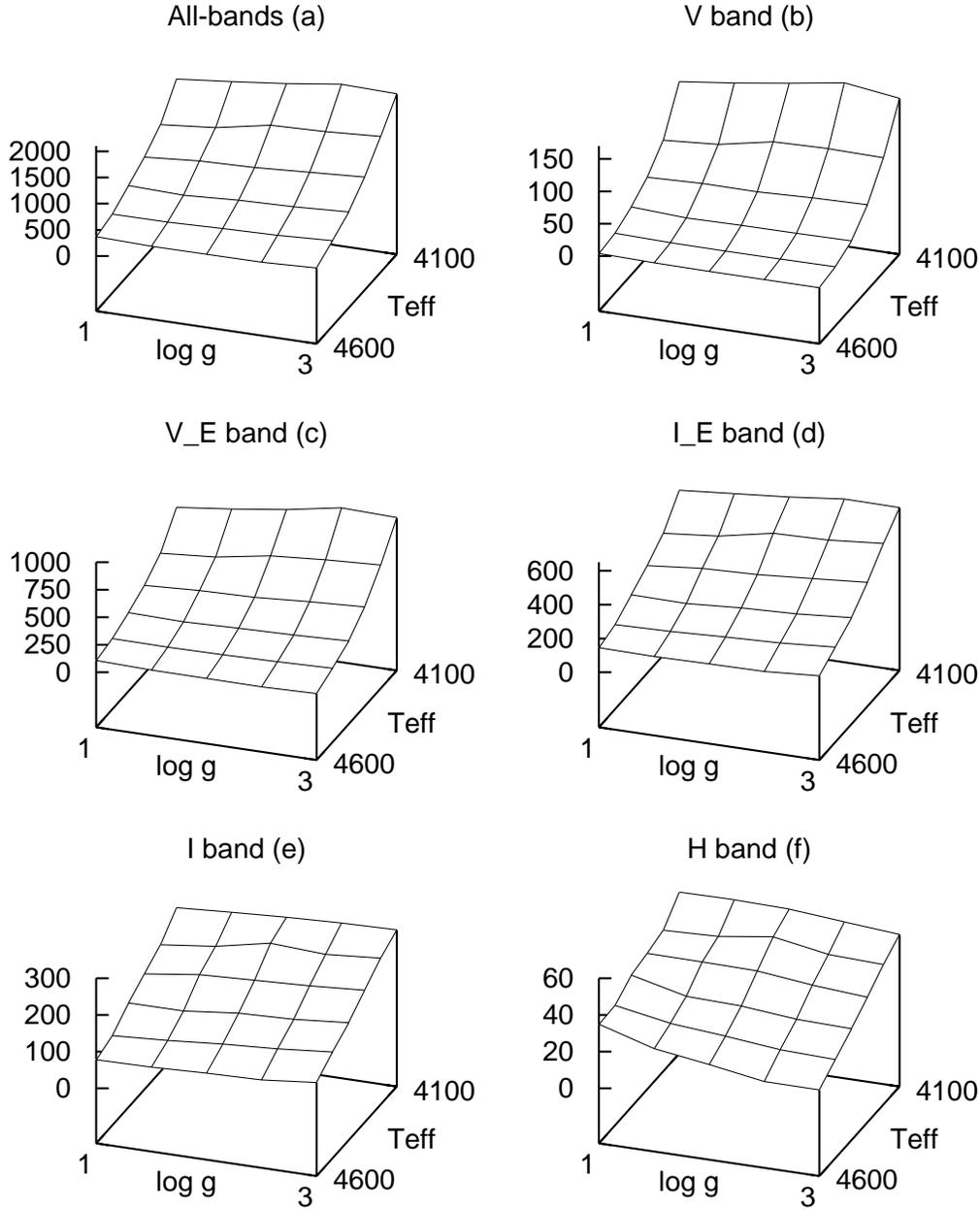}
\caption{\label{fig:chi2ng}
Shown are the $\chi^2$ surfaces of the microlens-Next$^2$Gen comparison
at [Fe/H]$=+0.3$ for (a) all the bands, (b) the $V$ band, (c) the $V_E$
band, (d) the $I_E$ band, (e) the $I$ band and (f) the $H$ band.  All
the surfaces share the same general shape: monotonic dependence on
$\Teff$ and [Fe/H], with a favored value of $\log g$.  With the
exception of very low $\log g$ (0.0-0.5), $\chi^2$ varies most in
$\Teff$, with variations over [Fe/H] and $\log g$ being much lower.
}\end{figure}

Turning now to the $\chi^2$ analysis, the MPM is immediately ruled out:
$\chi^2=1473$.  The PRMs are also ruled out, as the lowest $\chi^2$
among them is 837.  When we look over the entire parameter space, we
find $\chimin=227$ at $\Teff=4600\Kel$, $\log g=2.5$ and [Fe/H]$=-0.25$.
From Figure~\ref{fig:chi2ng}(a), it appears likely that the true
$\chimin$ is outside of the explored parameter space, at least in
$\Teff$.  There also appears to be a trend towards lower metallicity, so
it is possible that the best-fit metallicity is also outside our
explored parameter space.  We will discuss this further in
\S\ref{sec:sources}.  In general, $\chi^2$ increases with [Fe/H], and
decreases with $\Teff$.  Surface gravities between $\log g=1.0$ and
$2.5$ are preferred, with higher $\Teff$ and lower metallicity selecting
for a higher $\log g$.


\subsubsection{Johnson-Cousins $V$, $I$ and $H$ bands} \label{sec:ngvih}

At the MPM, the $V$, $I$ and $H$ bands have $\chi^2=93$, $237$ and $47$,
respectively.  Among the PRMs, all three bands prefer $\Teff=4300\Kel$
and [Fe/H]$=0.0$, though the $V$ and $H$ bands slightly prefer $\log
g=2.5$ while $I$ band slightly prefers $\log g=2.0$.  The $V$, $I$ and
$H$ bands have $\chi^2$ at these locations of 32, 172 and 40.  The
global minima for these bands are: $V$ band) $\Teff=4600\Kel$, $\log
g=3.5$ and [Fe/H]$=0.0$ with $\chimin=0.97$; $I$ band) $\Teff=4600\Kel$,
$\log g=2.5$ and [Fe/H]$=-0.25$ with $\chimin=60$; and $H$ band)
$\Teff=4600\Kel$, $\log g=3.0$, [Fe/H]$=-0.25$ and $\chimin=10.3$.  In
sum, the all-bands and $I$-band minima coincide, and the $\chi^2$'s for
bands $V$ and $H$ at the all-bands minimum is just $\dchi\sim2$ higher
than at their own minima.  That is, all three bands have the same
minimum to within $\sim1$ sigma.

Qualitatively, this shape of the $\chi^2$ surface is replicated for each
individual band as can be seen from Figure~\ref{fig:chi2ng}.  The minor
differences are that the $V$ band surface has more curvature in $\Teff$,
and $H$ band has more structure in $\log g$.



\subsubsection{EROS bands $V_E$ and $I_E$} \label{sec:ngeros}

These two bands are in greater disagreement with the atmospheric models
than any Johnson-Cousin band, except for $\Teff=4600\Kel$ and $\log
g\geq3.0$ ($\log g\geq2.0$ for [Fe/H]$=-0.25$).  In most cases this
disagreement is much greater.  The MPM has $\chi^2=593$ and $503$ for
$V_E$ and $I_E$, respectively.  The best PRM for both bands is
$\Teff=4300\Kel$, $\log g=2.0$ and [Fe/H]$=0.0$ with $\chi^2=287$ and
$302$.  The location of the $\chimin$ for the $V_E$ and $I_E$ bands is
the same as the all-bands, at $\Teff=4600\Kel$, $\log g=2.5$ and
[Fe/H]$=-0.25$ with $\chi^2=54$ and $99$.

\subsection{Possible Systematic Effects} \label{sec:sources}

Logically, there are only four possible sources for the discrepancy
between the models and the data: 1) problems with the microlensing data, 2)
problems with our analysis of the data, 3) problems with the atmosphere
models, or 4) incorrect comparison of the models and the data.  We now
argue that (1), (2) and (4) are unlikely.

\subsubsection{Individual Observatories} \label{sec:indobs}

We test whether the data from an individual observatory drives the
combined solution to an unsuitable answer.  We rerun our fitting routine
five times at the ($d$,$q$) of the combined-bands solution, each time
removing a different observatory's data, the exception being that we
always keep both $H$ band data sets.  The removal of the SAAO, Perth or
EROS data sets do not appreciably change the limb-darkening curves.  In
the $V$ band, the removal of the Canopus data shifts the limb-darkening
parameters by approximately $1\,\sigma$, and the removal of the YALO
data shifts them by about $3\,\sigma$.  The reverse is true in the $I$
band, removing the Canopus data provokes a $3\,\sigma$ change while
removing the YALO data only induces a $1\,\sigma$ change.  This test
shows that any systematic effects in the data themselves are either
present in data sets across all observatories, or are so mild that they
do not affect the combined solution at the level of the difference
between microlensing measurements and the atmospheric models.

\subsubsection{PLANET vs. EROS Datasets} \label{sec:pe}

We also analyze the solutions found by the PLANET data and EROS data
separately.  We expand the analysis to include searching for a solution
over $d$ and $q$.  The PLANET-only solution is located at the same
($d$,$q$) as the combined solution.  Moreover, as discussed in
\S\ref{sec:indobs}, removing the EROS data does not appreciably
change the limb-darkening profiles found by the remaining (i.e. PLANET)
data.  However, we find a very different result for the EROS-only
solution.  This is located at a ($d$,$q$) of (1.94,0.77) that is
(0.005,0.02) away from the combined solution.  The EROS solution has a
$\chi^2$ 75 lower than the $\chi^2$ of the EROS bands at the combined
solution.  The derived stellar brightness profile is flat across the
inner half of the star, then drops dramatically towards the limb.  The
ratio of intensities of the center to the limb are similar to that of
the $V$ band, but the shape of the EROS profiles are very different.
The EROS-only profiles are a much better match to the Next$^2$Gen models
than the EROS profiles derived at the all-bands geometry.  The $\chi^2$
of the MPM drops from 593 and 503 to 174 and 69 for the $V_E$ and $I_E$
bands, respectively.  However, the profiles from the EROS-only solution
are still not actually consistent with any of the Next$^2$Gen models, since
$\chimin=87$ and 32 in the two EROS bands.

Such a major inconsistency is a potentially serious problem.  How can
the previously described disagreement between atmospheric models and
microlensing be trusted if the microlensing can produce such different
fits?  We argue that this problem can be resolved in the following
manner: 1) we identify the feature that has the most diagnostic power
with regards to the limb darkening: the caustic exit, 2) we show how the
EROS datasets do not well determine this feature, though the combination
of EROS and SAAO $H$ band datasets do, and 3) we investigate whether the
large formal difference in $\chi^2$ between the EROS and all-bands
geometries represent a failing of the model or of the data, and what the
consequences of that failure are.

\begin{figure}
\plotone{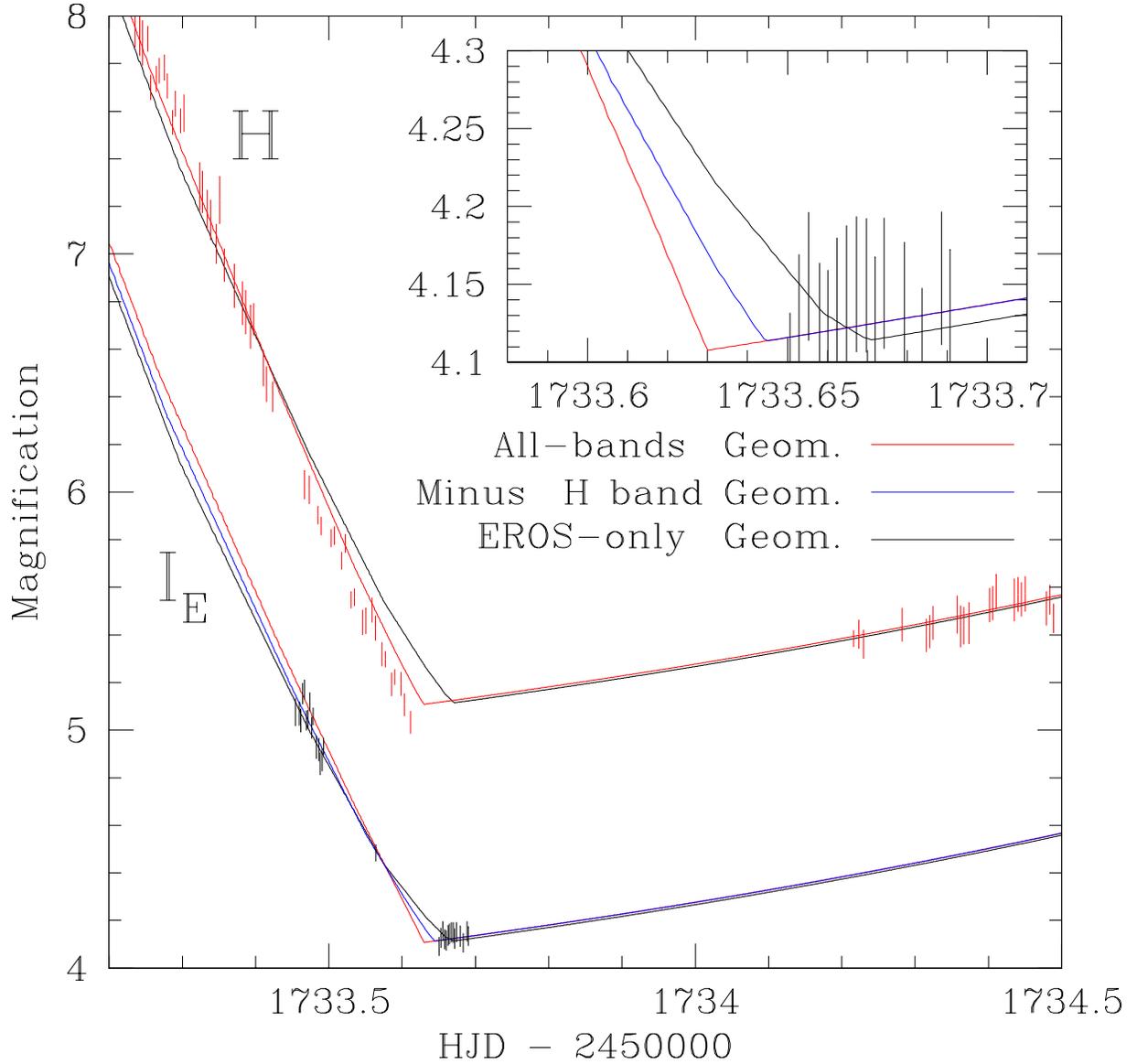}
\caption{\label{fig:lc}
The magnification for \eb\ in the $H$ and $I_E$ bands.  The $H$ band has
been shifted by $+1$ in magnification to separate it from the $I_E$
band.  Also shown are the error bars for the SAAO $H$ band and EROS
$I_E$ band data points.  The magnification for each data point has been
reconstructed using the observed flux and the source flux, blend flux
and seeing correction, the last three of which are derived from the
all-bands solution.  These three quantities are very stable and do not
appreciably change between microlens models.  The black lines show the
predicted light curve derived from the geometry at the EROS-only
solution, the blue line shows the predicted light curve from the
geometry at the solution containing all data sets except the $H$ band,
and the red lines show the predicted light curve for the geometry at the
all-bands solution.  The inset shows an expanded view of the $I_E$
caustic-exit region.  Note that the $I_E$ band points could support
either prediction, while the $H$ band points strongly favor the
all-bands solution.  }\end{figure}

We first examine the region from which we receive the most information
about the limb darkening: near the caustic exit.  This region is shown
in Figure~\ref{fig:lc}.  An accurate estimate of the caustic-exit time
is essential for determining the amount of darkening on the extreme
limb.  This can be illustrated by thinking of data points just outside
of the true caustic exit.  If the caustic exit were thought to occur
later than it actually does, these points would be thought to be inside
of the caustic.  Their faintness would therefore imply that the part of
the star undergoing the strongest differential magnification (the limb)
had very low surface brightness.  On the other hand, if the caustic exit
is recognized to occur before these points, their faintness is properly
attributed to the fact that there are no additional images of the
source, i.e. the source lies entirely outside the caustic.

This is exactly the issue with respect to the disagreement between the
PLANET-only and EROS-only geometries.  As Figure~\ref{fig:lc} shows, the
EROS-only geometry places the caustic exit at a later time, which
implies greater limb darkening.  We therefore investigate how well the
caustic-exit time is determined by the all-bands data set and what the
source of this discrepancy is.

First we note that the time of the caustic exit is essentially
determined from the combination of the SAAO $H$ band data and the EROS
data: the $H$ band data show an approximately linear fall toward the
caustic exit, and this fall must break very close to the best-fit
caustic exit if the magnification curve is to remain continuous and
still pass through the EROS $I_E$ points.  See Figure~\ref{fig:lc}.
Thus, the caustic-exit time can be specified virtually without reference
to any model.

Even if the EROS points are eliminated from the fit, the best-fit light
curve based on PLANET-only data still passes through these EROS points
and intersects the linearly-falling SAAO $H$ band lightcurve at almost
exactly the same caustic exit.  Because the PLANET points that fix the
post-exit magnification start up about 10 hours after the exit when
the magnification has already started to rise, this determination of the
caustic exit is somewhat model-dependent.  However, the model dependence
is quite weak.  Hence, we have two independent and robust lines of
evidence fixing the caustic-exit time, and for this reason we have high
confidence in the result.

Nevertheless, it remains somewhat puzzling why the EROS-only solution
prefers a later time.  From Figure~\ref{fig:lc}, it is clear that the
EROS data near the caustic exit do not themselves strongly prefer one
solution over the other.  Hence, this discrepancy must be rooted in
other parts of the light curve: either the EROS data have systematic
errors elsewhere in the light curve, or the model does not exactly
reproduce the true lightcurve.

To distinguish between these possibilities, we first exclude the $H$
band data and refit the lightcurve.  The result is shown by the blue
line in Figure~\ref{fig:lc}, which is between the EROS-only solution and
the PLANET-only solution.  Clearly the pressure toward a late caustic
exit is not coming from the EROS data alone.  To verify this, we
eliminate the $H$ band, the remaining SAAO bands, and the EROS data.
The resulting exit, which lies almost on top of the blue line of the
minus $H$ band geometry, also lies half way between the PLANET-only and
EROS-only values.  Since the problem is not restricted to one data set,
we conclude that the model must imperfectly predict the data elsewhere
in the lightcurve.  We have attempted to isolate this discrepancy using
various techniques, but have not succeeded because the effect is
extremely small and only manifests itself when data far from the caustic
exit are used to predict the caustic-exit time.  As with any such
extrapolation, and caustic-exit predictions in particular
\citep{complete}, small errors can be vastly magnified when predicting
distant effects.

We must also determine how much the fit to the caustic region (and so
the limb-darkening measurement) is being distorted on account of data
far away.  To do so, we decrease the error by a factor of ten on the
group of $H$ band points just before the caustic exit.  This should
increase the relative importance of this region to the overall fit and
be able to tell us to what extent the light curve near the caustic exit
is being influenced by data far away.  The caustic exit shifts slightly
to an earlier time, as we should expect given that data far from the
caustic tends to shift it to later times.  This shift, however, does not
produce substantial changes in the limb-darkening curves, shifting them
by 1-1.5$\,\sigma$ on average.  We conclude that the caustic exit is
very well determined, and the discrepancies related to it do not
significantly affect the limb-darkening determinations.

\subsubsection{Fitting Routine} \label{sec:routine}

It is unlikely that our fitting routine is the source of the conflict.
We fit all the data simultaneously, so one would expect any systematic
effects would have to be present in all bands.  The limb-darkening
curves in these five bands are all internally consistent with each
other.  Without outside information, the microlensing routine found that
the amount of limb darkening increases with decreasing wavelength,
starting from a very flat profile in $H$ band and progressing through
$I$, $I_E$, and $V_E$ to a relatively steep $V$ band profile (see
Fig.~\ref{fig:fivebands}).  Even the relative amounts of limb darkening
are roughly correct.  The mean wavelength of the $I_E$ band is $14\%$ of
the way between that of the $I$ and $V$ bands \citep{SMC}, and indeed,
the $I_E$ band limb-darkening profile is very similar to the $I$ band
profile.  The mean wavelength of the $V_E$ band is $27\%$ of the way
between $V$ and $I$ \citep{SMC}, and the $V_E$ profile is more like the
$V$ band profile than the $I$ band profile, as expected the degree of
similarity between the $V_E$ and $V$ profiles does not match that
between the $I_E$ and $I$.

\subsubsection{Atmospheric Models} \label{sec:atmo}

\begin{figure}
\plotone{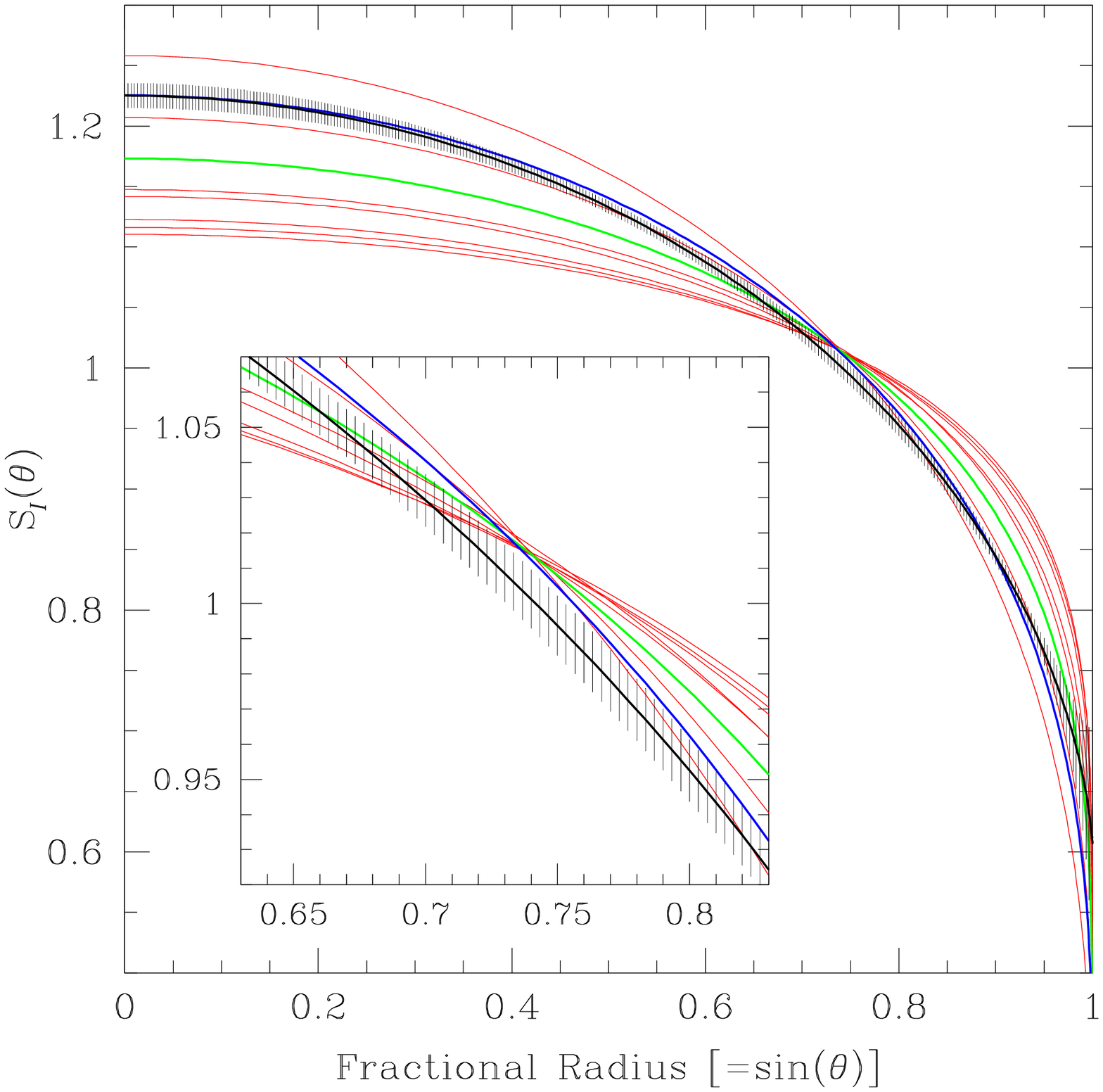}
\caption{\label{fig:ibandat}
The brightness profile in the $I$ band.  The black line and shading are
\eb\ and its associated $3\,\sigma$ error envelope.  The red lines give ATLAS
models \citep{Cl00} at $\log g=2.5$ and [M/H]$=+0.3$ for $\Teff$ between
3000 and 10000 K, in 1000 K increments.  The blue line corresponds to
the best-fit ATLAS model at $\Teff=4500\Kel$, $\log g=3.0$ and
[M/H]$=-0.3$.  Also shown is the $I$ band curve of the Sun in green.  As
can be seen, there is a point that all of the model curves share with
the Sun at approximately ($r$,$S_I$)=(0.74,1.02).  This fixed point
which is not shared by the microlensing-determined profile is not
apparent in the ($c_{\lambda}$,$d_{\lambda}$) formalism.  }\end{figure}

\begin{figure}
\plotone{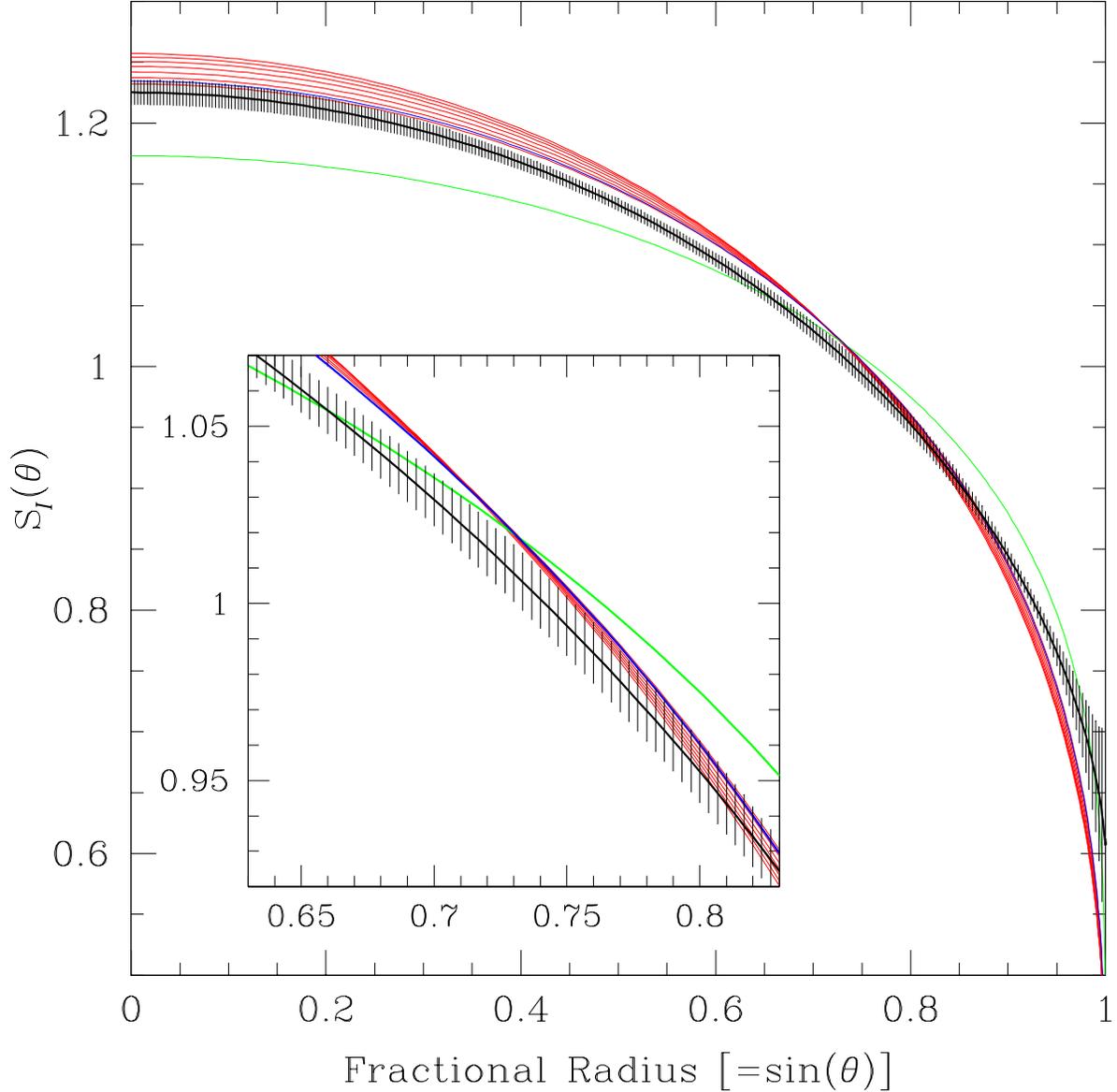}
\caption{\label{fig:ibandng}
The brightness profile in the $I$ band.  The black line and shading are
\eb\ and its associated $3\,\sigma$ error envelope.  The red lines
give Next$^2$Gen models at $\log g=2.5$ and [M/H]$=+0.3$ for $\Teff$
between 4100 and 4600 K, in 100 K increments.  The blue line corresponds
to the best-fit Next$^2$Gen model at $\Teff=4600\Kel$, $\log g=2.5$ and
[M/H]$=-0.25$.  Also given is the $I$ band curve of the Sun in green.
As can be seen, there is a point that all of the model curves share with
the Sun at approximately ($r$,$S_I$)=(0.73,1.02).  This fixed point is
the same as the one that the ATLAS models share (but the
microlensing-determined profiles do not) and is not apparent in the
($c_{\lambda}$,$d_{\lambda}$) formalism.  }\end{figure}

Figure~\ref{fig:ibandat} compares the $I$ band limb darkening derived
from microlensing to a suite of ATLAS atmospheric models all at the same
$\log g=2.5$ and [M/H]$=+0.3$, but with different temperatures.  A
striking feature of this figure is that all of the atmospheric models go
through a single point, one which the microlensing model does not go
through.  Note that the curve for the Sun (the only star in the sky with
better measured limb darkening than \eb) also passes through this point.
That is, the models all carry a common feature that is also present in
the Sun, but does not exist in the giant \eb.  This exact feature is
also present in the Next$^2$Gen models shown in
Figure~\ref{fig:ibandng}.  We conjecture that these models, which are
capable of producing fits to the Sun, incorporate physics that is only
applicable in the solar regime.  This common feature is not apparent
when the limb-darkening curves are normalized in the usual
($c_{\lambda}$,$d_{\lambda}$) formalism.  It appears only when
limb-darkening curves are plotted to conserve total flux as in
equation~(\ref{eqn:limblaw}).  In any event, this feature gives an
easily identifiable handle on the difference between the limb darkening
of the models and the limb darkening observed via microlensing.

\subsubsection{ATLAS-Next$^2$Gen Comparison} \label{sec:models}

The $\chi^2$ surfaces for the ATLAS and Next$^2$Gen models are shown in
Figures~\ref{fig:chi2at} and \ref{fig:chi2ng}.  As there are three input
parameters, $\Teff$, $\log g$ and metallicity for each model, we hold
the least important constant, and show $\chi^2$ versus the other two.
This results in two fairly similarly looking surfaces.  In both cases,
$\Teff$ is the most important parameter in determining $\chi^2$ with the
metallicity and surface gravity playing far more subordinate roles.
Additionally, while the true $\chimin$ for the Next$^2$Gen models is
beyond the range in parameter space we investigated, the flattening of
the $\chi^2$ surface indicates that it is not very far away.  Both the
ATLAS and the Next$^2$Gen models are most consistent with a star of
$\Teff\sim4700\Kel$, $\log g=2.5$-3.0 and metallicity around $-0.3$.

\subsubsection{Model Atmosphere-Microlens Profile Comparison} \label{sec:reddening}

As we have just shown in \S\ref{sec:models}, the atmospheric models both
seem to prefer a K1 star.  It is therefore important to ask if our
observations are biased and we are expecting the wrong type of star.  In
principle, differential extinction across the field could redden the
star more than the clump stars against which it is calibrated (see
Fig.10 of \citet{A02}), thus affecting our photometric estimate of its
intrinsic color.  However, the spectroscopic analysis yields a source
temperature similar to the one we find photometrically.

One should note that reddening has the effect of shifting the observed
bands.  We find, however, that the measured extinction for
\eb\ [$E$($V$-$I$)=1.3] shifts the mean wavelength of the $I$ bandpass
by only $+100$\AA.  This shift is about $28\%$ of the difference between
the mean wavelengths of $I$ and $I_E$ and only $4\%$ of the difference
between $I$ and $V$.  Comparing Figures~\ref{fig:ibandat} and
\ref{fig:fivebands}, one sees that it cannot account for differences
between the observations and the atmospheric models.  This is partly
because the magnitude of the effect is too small, and partly because it
causes a shift in the wrong direction.

\section{Conclusion} \label{sec:conclusions}

The observational limb darkening found from microlensing formally
disagrees with the limb darkening derived from atmospheric models by
many sigma.  We have argued that this difference is unlikely to be the
result of the observations, but is more likely due to something related
to the atmospheric models.  It is possible that these models include
physics that is not applicable in all surface-gravity regimes.  It is a
testament to the theoretical models that they approximate reality in
several bands without any previous physical data.  We hope that now that
giant stars have a calibration point, much like dwarfs have from the sun
and supergiants have from interferometry, stellar models can continue to
improve in all stellar regimes.

\begin{acknowledgements}

We would like to thank Joachim Wambsganss for his comments on this
manuscript.  DF and AG were supported by grant AST 02-01266 from the
NSF.  BG was supported by NASA through a Hubble Fellowship grant from
the Space Telescope Science Institute, which is operated by the
Association of Universities for Research in Astronomy, Inc., under NASA
contract NAS5-26555.

\end{acknowledgements}

\clearpage

\newpage
\begin{deluxetable}{cccc}
\tablecaption{Data Sets
\label{tab:data}}
\tablehead{
\colhead{Band}
&\colhead{Observatory}
&\colhead{N points}
&\colhead{$\sigma$ scaling}
}\startdata
$V$&SAAO&$177$&$1.47$\\
&Canopus&$154$&$1.96$\\
&YALO&$233$&$1.25$\\
$V_E$&EROS&$830$&$1.59$\\
$I_E$&EROS&$904$&$1.63$\\
$I$&SAAO&$404$&$1.79$\\
&Canopus&$311$&$2.57$\\
&YALO&$424$&$1.57$\\
&Perth&$148$&$2.89$\\
$H$&SAAO&$549$&$2.69$\\
&YALO&$659$&$1.95$\\
\enddata
\end{deluxetable}

\begin{deluxetable}{ccc}
\tablecaption{Model Parameters for \eb\
\label{tab:soln}}
\tablehead{\colhead{}&\colhead{Joint Solution}&\colhead{\citet{A02}}
}\startdata
$d$&$1.935$&$1.928\pm0.004$\\
$q$&$0.75$&$0.7485\pm0.0066$\\
$\alpha^\prime$&$73\fdg669054$&$74\fdg18\pm0\fdg41$\\
$u_{\rm c}$&$-5.110\times 10^{-3}$&$(-5.12\pm0.03)\times 10^{-3}$\\
$t_{\rm E}^\prime$&$100.030$ days&$99.8\pm1.5$ days\\
$t_{\rm c}$&$1736.941667$\,\tablenotemark{a}&$1736.944$\,\tablenotemark{a}\ $\pm0.005$\\
$\rho_\ast$&$4.792\times 10^{-3}$&$(4.80\pm0.04)\times 10^{-3}$\\
$\pi_{{\rm E},\parallel}$&$-0.166358$&$-0.165\pm0.042$\\
$\pi_{{\rm E},\perp}$&$0.195334$&$0.222\pm0.031$\\
$\dot d$&$0.283$ yr$^{-1}$&$0.203\pm0.016$ yr$^{-1}$\\
$\omega$&$0.0066$ rad yr$^{-1}$&$0.006\pm0.076$ rad yr$^{-1}$\\
\enddata
\tablenotetext{a}
{Heliocentric Julian Date $-$ 2450000.}
\end{deluxetable}

\newpage
\begin{deluxetable}{cccccc}
\tablecaption{Limb-Darkening Parameters, Errors and Correlation Coefficients
\label{tab:corr}}
\tablehead{
\colhead{}
&\colhead{$V$ band}
&\colhead{$V_E$ band}
&\colhead{$I_E$ band}
&\colhead{$I$ band}
&\colhead{$H$ band}
}\startdata
$\Gamma$&$0.7077$&$0.4813$&$0.6494$&$0.5173$&\nodata\\
$\sigma_{\Gamma}$&$0.1021$&$0.0855$&$0.0656$&$0.0726$&\nodata\\
$\Lambda$&$0.0456$&$0.2880$&$-0.3305$&$-0.1336$&$0.4782$\\
$\sigma_{\Lambda}$&$0.1871$&$0.1621$&$0.1264$&$0.1360$&$0.0154$\\
$\tilde{c}_{ij}$&$-0.9966$&$-0.9979$&$-0.9978$&$-0.9974$&\nodata\\
\enddata
\end{deluxetable}

\end{document}